\documentstyle[preprint,aps]{revtex}
\begin{document}

\title{Folding and Aggregation of Designed Proteins}

\author{R.A. Broglia$^{1,2}$, G. Tiana$^3$, S. Pasquali$^1$,  H.E. Roman$^{1}$
        and E. Vigezzi $^{1}$}

\address{$^1$Dipartimento di Fisica Universit\`a di Milano and I.N.F.N.,  
             I-20133~Milano,~Italy.}

\address{$^2$The Niels Bohr Institute, University of Copenhagen, 2100 Copenhagen,
             Denmark.}

\address{$^3$Department of Physics, DTU Building 307, 2800 Lyngby, Denmark.}


\maketitle

\narrowtext 
\begin{abstract} 
{\it Studies of how protein fold have shown that the
way protein clumps form in the test tube is similar to how proteins form the
so-called ``amyloid'' deposits that are the pathological signal of a variety of
diseases, among them the memory disorder Alzheimer's
\cite{Fink98,Silow,Mitraki,Wetzel98,Janicke,Wetzel94}. Protein 
aggregation have traditionally been connected to either unfolded or native states. 
Inclusion body
formation (disordered aggregation) has been assumed to arise from hydrophobic
aggregation of the unfolded or denaturated states, while the amyloid fibrils
(ordered aggregation) have been assumed to arise from native-like conformations in
a process analogous to the polymerization of hemoglobin S. Making use of
lattice-model simulations \cite{Shakh96,Shakh94,Tiana} we find that both ordered
and disordered aggregation arise from elementary structures which eventually
build the folding nucleus of the heteropolymers,
and takes place when some of the most strongly interacting amino acids establish
their contacts leading to the formation of a specific subset of
the native structure. These elementary structures can be viewed as the partially 
folded intermediates suggested to be involved in the aggregation of a number of 
proteins \cite{Wetzel94,King,Speed,Hurla,Kim,Fink95}. These results have 
evolutionary implications, as the elementary structures forming the
folding core of designed proteins contain the residues which
are conserved among the members of homologous sequences.}
\end{abstract}

\bigskip
\bigskip
\bigskip
\bigskip

\narrowtext 

There are still many outstanding and critical questions
regarding protein aggregation, despite many studies devoted to the subject. Among
these are questions concerning the detailed mechanism of the aggregation process.
We approach this problem within the framework of a simple lattice model of protein 
folding \cite{Shakh96,Shakh94,Tiana} and study, making use of Monte Carlo (MC) 
simulations, the simultaneous folding of
two identical twenty-letter amino acid chains each composed of 36 monomers, and
designed to fold into their native conformation (Figs.~1(a) and 1(b)).

Three different outcomes of the simulations have been observed: (I) both chains
fold  to their native conformation   (Fig.~1(a)), (II) one of the chains folds while
the other attaches to it in a compact configuration (Fig.~1(c)), (III) both chains
get deeply intertwined in conformations which are quite compact and display some
amount of similarity to the native conformation (Fig.~2).  Situations
(II) and (III) are typical of ordered and disordered aggregation, respectively.
In case (I) each chain targets into its minimum energy structure (native
conformation) about which it fluctuates \cite{Ptitsym,Shakh89}. Cases (II) and
(III) are associated with an ensemble of compact low-energy conformations typical of
those reached in the folding of a random chain, where the system spends little time
in each conformation and displays conspicuous energy fluctuations.

At the basis of these phenomena are the elementary structures built out of the
monomer sequences S$^1_4\equiv(3,4,5,6)$, S$^2_4\equiv(27,28,29,30)$, and
S$^3_4\equiv(11,12,13,14)$ (cf. Fig.~1(a)), containing essentially all of the
amino acids forming the folding nucleus \cite{Shakh96} of the designed sequence
S$_{36}$ (Fig.~1(b)). In fact, the structures S$^i_4$ ($i=1,2,3$) can be
viewed as the (dynamical) ``bricks'' of a LEGO kit to model proteins.

The pairs of monomers (3,6), (27,30) and (11,14) become nearest neighbours very early
in the folding process, the associated first passage time (FPT) being 104, 102 and 
260 MC steps, respectively. The corresponding contacts achieve 90-95\% stability
already after $0.25\times 10^6$ MC steps, a time to be compared with the FPT for the 
folding of both interacting chains (situation (I)) and equal to $2\times 10^6$ MC steps.
The folding core is formed essentially when the three different ``bricks'' of the same 
chain assemble together, establishing the contacts 6-27, 3-30, 6-11 and 27-14, at which 
time it becomes easy for the contacts 27-16 and 30-33 to fall in place. Once the folding
nucleus of both proteins is formed, it takes less than $3\times 10^4$ MC steps for them to 
reach the native configuration. All the contacts which maintain the ``bricks'' in place 
involve at least one amino acid occupying a ``hot'' site in the native conformation
of the isolated protein \cite{Tiana}, that is a strongly interacting amino acid
(Fig.~1(a)). In the situation under study, there are 3 ``hot'' sites, namely
sites number 6, 27 and 30. Once the "hot" site amino acids are in place, it takes
$0.6\times 10^6$ MC steps for both proteins to fold (FPT), in keeping with the
fact that while the FPT of the contact 6-27 is $\approx 0.4\times 10^6$ MC steps,
it takes $\approx 1.4\times 10^6$ MC steps for it to become stable.

Aggregation results because of the exchange taking place between the interacting
chains, of the role played by amino acids occupying ``hot'' sites, a phenomenon
whose associated FPT is typically $0.5\times 10^6$ MC steps. In other words,
aggregation happens when ``bricks'' belonging to different chains attach to each
other (cf. Fig.~1(c) and Fig.~2). Such a ``mistake'' can happen in a number
of different ways, and not only in the one which mimicks the disposition of the
``bricks'' in the native core configuration, in keeping with the LEGO analogy.
Because of the strongly interacting character of the amino acids occupying sites 27, 30 and 6,
aggregation is, for all purposes, an irreversible process under native like conditions,
as testified by the results of simulations leading to aggregtion which have been followed
over $10^8$ MC steps.

The rate of aggregation is found to depend sensitively on protein concentration, in
keeping with the fact that the likelihood that the elementary structures belonging
to different chains interact depends on the average distance between the 
heteropolymers. A number of observations testify to the
central role of protein concentration on the phenomenon of aggregation 
\cite{London,Kelley,Citron,Taubes,Zettmeissl,Vaucheret,Garel}. In Fig.~3 we display
the calculated aggregation probability as a function of concentration, in comparison 
with the results of observations \cite{Zettmeissl,Vaucheret}. Theory provides an 
overall account of the experimental findings.

We have found that the rate of aggregation increases in a significant manner, by
introducing ``cold'' (neutral) mutations \cite{Tiana}. The chosen mutations are able to 
affect in a significant way the stability of one of the elementary structures, without
much changing the ability the resulting isolated sequence S$'_{36}$ have to fold on short 
call to the native conformation. In particular, by substituting the amino acid $R$ at
position 11 of the designed sequence, by amino acid $A$, the rate with which
aggregation of type (III) takes place increases by 70\% (i.e. from  22\% to a 37\%
rate). The reason for this increase lies in the fact that for the resulting sequence S$'_{36}$
it takes $0.6\times 10^6$ MC steps for the pairs (11,14) to become nearest
neighbours (as compared to $0.25\times 10^6$ MC steps for S$_{36}$). Consequently,
the other two elementary structures (associated with the monomer groups S$^1_4$ and 
S$^2_4$) have more time and 
thus a better chance to interact with the homologous structures of the other chain, 
than in the case of the simultaneous folding of two S$_{36}$ sequences.
Similar results have been obtained by performing single and multiple 
mutations in ``cold'' and ``warm'' sites of the native conformation. 
Because 75\% of all sites are ``cold'' sites, and thus
associated with neutral mutations, there is a large number of mutations
which, while 
destabilizing the elementary structures, and thus increasing the rate of
aggregation, do not affect in an important way the stability of the protein. 
These results are
consistent with a number of observations, in particular those carried out in the
study of the amyloid-forming system, transthyretin. When altered by 
any of 50
different mutations, this protein, which normally occurs in the blood plasma,
deposits in the heart, lungs and gut, causing a lethal disease called familial
amyloidotic polyneuropathy (FAP)   \cite{McCutchen93,McCutchen95}. These mutations
do not alter normal folding of the protein but do
destabilize the protein structure, facilitating the formation of partially folded
intermediates that readily aggregate to one another \cite{Hamilton,Terry}.

We conclude that a given protein will have a very small number of
partially folded intermediates which controls both protein folding and aggregation. 
Within the model of designed proteins these are the elementary structures
which build the folding nucleus. Consequently,  most of the aggregates of this
protein as well as of the sequences homologous to it, will display similar
native-like structures, independent of the nature of the effect triggering the
aggregation.

\bigskip 
\bigskip 
\bigskip 
\bigskip 
\bigskip 

\newpage
\leftline{ACKNOWLEDGMENTS}

We gratefully acknowledge financial support by NATO under grant CRG 940231.
 We also thank the late Dr. N. D'Alessandro for help in modelling design.
The colour plots in Figs.~1 and 2 have been obtained with the graphic program of 
Ref.\cite{VMD}.


\newpage


\centerline{\bf Fig.1}
\noindent
         {\bf a}, The conformation of the 36-mer chosen as the native state in 
         the design procedure. Each amino-acid residue is represented as a bead
         occupying a lattice site. Although the model does not treat side chains
         explicitely, the amino acids are chemically different. Their differences
         are manifested in pairwise interactions energies of different magnitude
         and sign, depending on the identity of the interacting amino acid. The
         configurational energy is
         $$E=\protect\frac{1}{2}\protect\sum_{i,j}^{N}U_{m(i),m^{\protect\prime }(j)}~ 
         \protect\Delta (|\protect\stackrel{\protect\rightarrow }{r_i}-
         \protect\stackrel{\protect\rightarrow }{r_j}|),$$
         $\{ \protect\stackrel{\protect\rightarrow}{r} \}$ being the set of 
         coordinates of all the monomers describing a chain conformation. The quantity
         $\protect\Delta (|\protect\stackrel{\protect\rightarrow }{r_i}-
          \protect\stackrel{\protect\rightarrow }{r_j}|)$ is a contact function.
          It is equal to one if sites $i$ and $j$ are at unit distance (lattice
          neighbors) not connected by a covalent bond, and zero otherwise. In addition,
          it is assumed that on-site repulsive forces prevent two amino acids to
          occupy the same site simultaneously, so that $\Delta(0)=\infty$. There
          are 20 types of amino acids in the model. The quantities 
          $U_{m(i),m^{\protect\prime }(j)}$ are the contact energies
          between amino acids of type $m$ and $m'$, and were taken from Table 6 of
          ref.\cite{Miyazawa}. The 36-mer chain denoted S$_{36}$ and designed by
          minimizing, for fixed amino acid concentration, the energy of the native 
	  conformation with respect to the amino acid sequence,
          folds in $10^6$ MC steps at the optimal temperature $T=0.28$ (in our
          temperature scale). The effects mutations (19 possible substitutions of
          monomers on each site) have on the folding properties of S$_{36}$ have been
          studied in \protect\cite{Tiana}. It was found that the 36 sites of the
          native conformation can be classified as ``hot'' (red beads, numbered 6, 27 
          and 30), ``warm'' (beads numbered 3, 5, 11, 14, 16 and 28) and ``cold'' 
          (the rest of the beads) sites. In average,
          mutations on the 27 cold sites yield sequences which still
          fold to the native structure (neutral mutations), although the folding
          time is somewhat longer than for S$_{36}$. Sequences obtained from mutations
          on the 6 ``warm'' sites
          fold, as a rule, to a unique conformation, sometimes different but in any 
          case very similar to the native one. 
          Mutations on the three "hot" sites lead, in average, to complete
          misfolding (denaturation) of the protein. The design tend to place the most
          strongly interacting amino acids in the interior of the protein where they can
          form most contacts. The strongest interactions are between groups
          $D, E$ and $K$ (cf. {\bf b}), the last one being buried deep in the protein 
	  (amino acid
          in site 27). The folding nucleus (determined from the folding simulations) is
          formed by these amino acids (red beads) and by their nearest-neighbours 
	  (yellow beads), and is shown by continuous blue lines. The structures 
	  formed by the amino acid sequences
	  S$^1_4\equiv(3,4,5,6)$, S$^2_4\equiv(27,28,29,30)$ and
          S$^3_4\equiv(11,12,13,14)$ of chain 1 are explicitely shown making use of 
	  light violet, light red and light yellow shades.	 	  	  
          {\bf b}, Designed amino acid sequence S$_{36}$.
          {\bf c}, Example of ordered aggregation. Chain 2  has folded to its native
	  conformation, while chain 1 has become attached to it. 
          The hot sites of chain 2 are shown as blue beads, the corresponding
          nearest neighbours amino acids in the native conformation by green beads.
	  The three basic structures of chain 2 (left), are shown in terms of light
	  green, green and light blue shades.
          The ``correct'' interactions of the folding nucleus of chain 1 are shown 
	  with a continuous line. The ``wrong'' ones, by dashed lines.
          The interactions associated with the folding core of the configuration
	  associated with chain 2 have not been shown, as they are identical to
	  those displayed for chain 1 in {\bf a} making use of continuous blue lines.


\bigskip
\bigskip
\bigskip
\bigskip

\centerline{\bf Fig.2}
\noindent
         Examples of disordered aggregation, where none of the chains have folded,
	 but have intertwined. 
	 The presence of yellow beads close to the blue ones, and of green beads close
	 to red ones, as well as of dashed
         lines connecting nearest neighbours, indicate that chain 1 has
         erroneously interpreted some of the elementary structures of chain 2 as
	 belonging to itself and viceversa. 	 
	 {\bf a}, In this case, aggregation is controlled by the
	 elementary structures made out of the sequences S$^1_4$ and S$^2_4$.
	 {\bf b}, Example of aggregation controlled by the elementary structures
	 built out of sequences S$^2_4$ and S$^3_4$.

\bigskip
\bigskip
\bigskip
\bigskip
\newpage

\centerline{\bf Fig.3}
\noindent
         Folding probability of the lactate dehydrogenase protein (LDH) as 
         a function of protein concentration (shaded area) taken from refs.
         \protect\cite{Zettmeissl,Vaucheret,Garel}. The width of the shaded area 
         takes into account the dispersion of the experimental values.
         The results of the model calculations are 
         displayed by solid dots.
         An initial configuration of two S$_{36}$-chains was
         generated by first locating 
         the 18th monomer (center) of one chain at the origin
         of coordinates, followed 
         by the growth of a self-avoiding random walk for its
         remaining 35 monomers. 
         Then, the second chain, which does not intersect
         neither itself nor the 
         previously created chain, was generated similarly by
         locating its 18th monomer 
         at a distance $d$ from the center of the first
         chain. The mean radii $\sigma$ 
         of each chain configuration were found to be the
         same, independently of $d$,
         and given by $\sigma\protect\cong 3.1$.
         The average distance $d_c$ between the two
         heteropolymer chains is defined 
         as the distance between their centers of mass. As a
         function of $d$, the
         following values were obtained: $d_c=$3.2$\pm0.1$,
         3.6$\pm0.1$, 4.9$\pm0.1$, 
         6.2$\pm0.1$ and 10$\pm0.1$ for $d=\sqrt{2}$, 2, 4, 6
         and 10, respectively.
         In this case, the ``concentration'' $c(d_c)$ of
         chains can be estimated as 
         $c\protect\cong c_0 / d_c^3$, where the normalization constant $c_0$ 
         has been set $c_0=13$ nM.


\begin{thebibliography}{Dillo 99}

\bibitem{Fink98} A.L. Fink, {\it Folding and Design} {\bf 3}, R9-R23 (1998) 

\bibitem{Silow} M. Silow and M. Oliberg, 
                {\it Proc. Natl. Acad. Sci. USA } {\bf 94}, 6084  (1997)

\bibitem{Mitraki} A. Mitraki and J. King, 
                  {\it Biotechnology} {\bf 7}, 690 (1989)

\bibitem{Wetzel98} R. Wetzel, {\it Cell} {\bf 86}, 699 (1998)

\bibitem{Janicke} R. Janicke, 
                  {\it Phil. Trans. Roy. Soc. London} {\bf B 348}, 97 (1995)

\bibitem{Wetzel94} R. Wetzel, 
                   {\it Trends in Biotechnology} {\bf 12}, 193 (1994)

\bibitem{Shakh96} E.I. Shakhnovich, V. Abkevich and O. Ptitsym, 
                  {\it Nature} {\bf 379},  96 (1996)

\bibitem{Shakh94} E.I. Shakhnovich, 
                  {\it Phys. Rev. Lett.} {\bf 72}, 3907 (1994)

\bibitem{Tiana} G. Tiana, R.A. Broglia, H.E. Roman, E. Vigezzi and E.I. Shakhnovich, 
                {\it J. Chem. Phys.} {\bf 108}, 757 (1998)

\bibitem{King} J. King, C. Haase-Petingell, A.S. Robinson, M. Speed and A. Mitraki,
               {\it FASEB J.} {\bf 10}, 57 (1996)

\bibitem{Speed} M.A. Speed, T. Morshead, D.I.O. Wang and J. King,
                {\it Protein Science} {\bf 6}, 99  (1997)

\bibitem{Hurla} M.R. Hurla, L.R. Helms, L. Li, W. Chan and R. Wetzel,
                {\it Proc. Natl. Acad. Sci.  USA } {\bf 91}, 5446  (1994)

\bibitem{Kim} D. Kim and M.H. Yu, 
              {\it Biochem. Biophys. Res. Comm.} {\bf 226}, 378 (1996) 

\bibitem{Fink95} A.L. Fink, 
                 {\it Ann. Rev. Biophys. Biomol. Struct.} {\bf 24}, 495 (1995). 

\bibitem{Ptitsym} O.B. Ptitsym, 
                  in {\it Protein Folding} (Freeman, New York 1992), pp. 243-300.

\bibitem{Shakh89} E.I. Shakhnovich and A.V.  Finkelstein, 
                  {\it Biopolymers} {\bf 28}, 1667 (1989)

\bibitem{London} J. London, C. Skrzynia and M.E. Goldberg, 
                 {\it European J. Biochem.} {\bf 47}, 409 (1974)
                                                  
\bibitem{Kelley} J.W. Kelley, 
                 {\it Current Opinion in Structural Biology} {\bf 6}, 11 (1996)                      
  

\bibitem{Citron} M. Citron, T. Oltersdorf, C, Haass, L. McConlogue, A. Y. Hung, 
                 P. Senbert, C. Vijo-Pelfrey,  I. Lieberburg and D. Selkoe, 
                 {\it Nature} {\bf 360}, 672 (1992)

\bibitem{Taubes} G. Taubes, 
                 {\it Science} {\bf 271}, 1493 (1996).
                
\bibitem{Zettmeissl} G. Zettmeissl, R. Rudolph and R. Jaenicke,  
                     {\it Biochemistry} {\bf 18},  5567 (1979)
                     
\bibitem{Vaucheret}  H. Vaucheret, L. Signon, G. Le Bras and J.-R. Garel,
                     {\it Biochemistry} {\bf 26}, 2785 (1987).

\bibitem{Garel} J.R. Garel, in {\it Protein Folding},  Ed. E. Creighton 
                (W.H. Freeman, New  York 1992), p.405.

\bibitem{McCutchen93} S.L. McCutehen, W. Colon and J.W.  Kelley,
                      {\it Biochemistry} {\bf 32}, 12119 (1993)

\bibitem{McCutchen95} S.L. McCutehen, Z. Lai, G. Miroy, J.W.  Kelley and W. Colon,
                      {\it Biochemistry} {\bf 34}, 13527 (1995)

\bibitem{Hamilton} J.A. Hamilton, L.K. Steinraut, B.C. Braden, J. Liepnieks, 
                   M.D. Benson, G. Holmgren, O. Sandgren and L. Steen, 
                   {\it J. Biol. Chem.} {\bf 268}, 2425 (1993)

\bibitem{Terry} C.J. Terry, A.M. Damas, P. Oliveira, M.J. Saraivia, I.L. Alves, 
                P.P. Costa, P.M. Matias, Y. Sakaki and C.C.F. Blake, 
                {\it EMBO J.} {\bf 12}, 735 (1993)

\bibitem{Miyazawa} S. Miyazawa and R.L. Jerningan,
                   {\it Macromolecules} {\bf 18}, 534 (1985)

\bibitem{VMD} W. Humphrey, A. Dalke and K. Schulten,   
              {\it J. Mol. Graphics} {\bf 14}, 33 (1996).

\end{thebibliography}
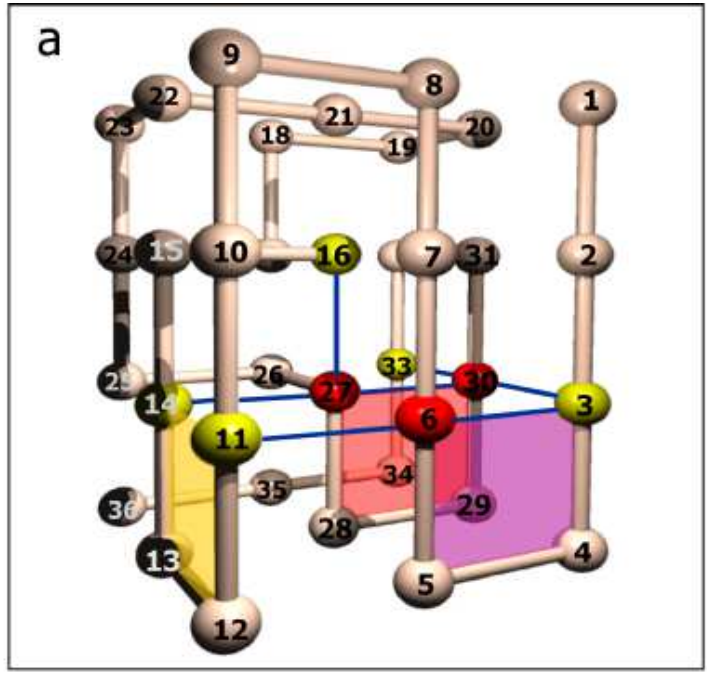
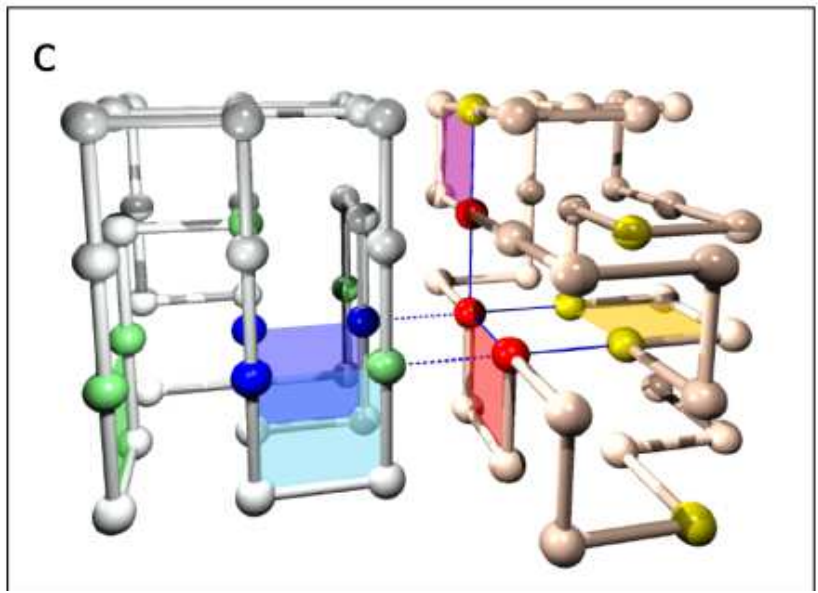
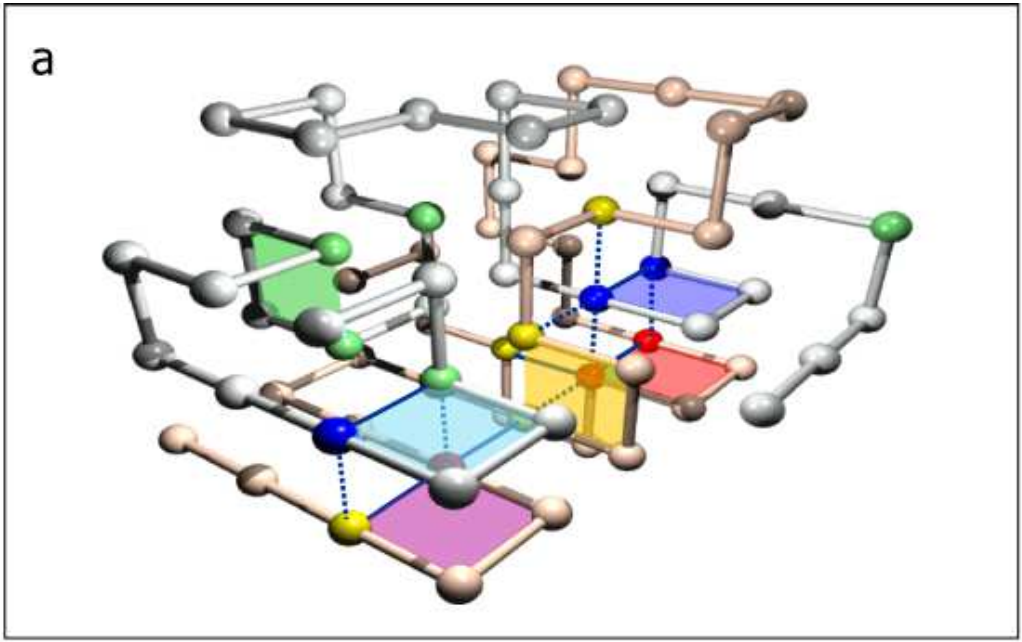
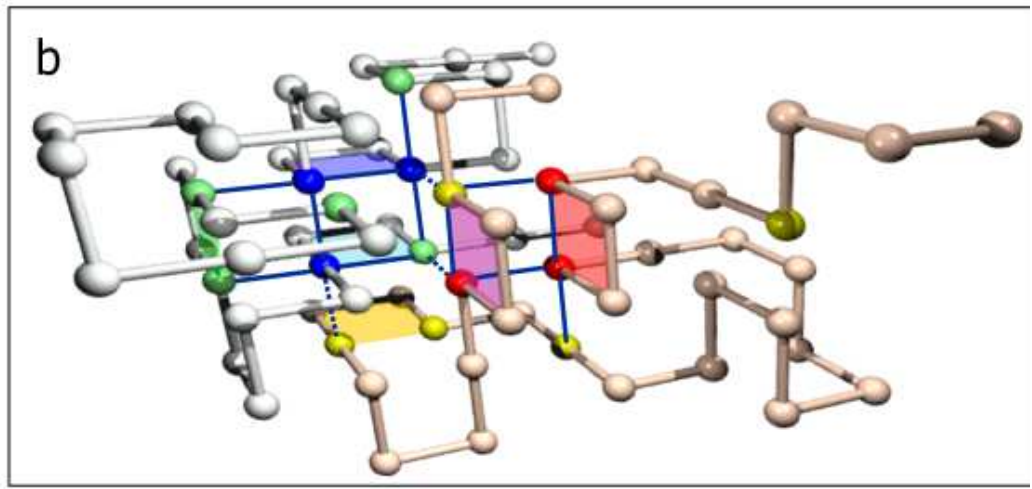
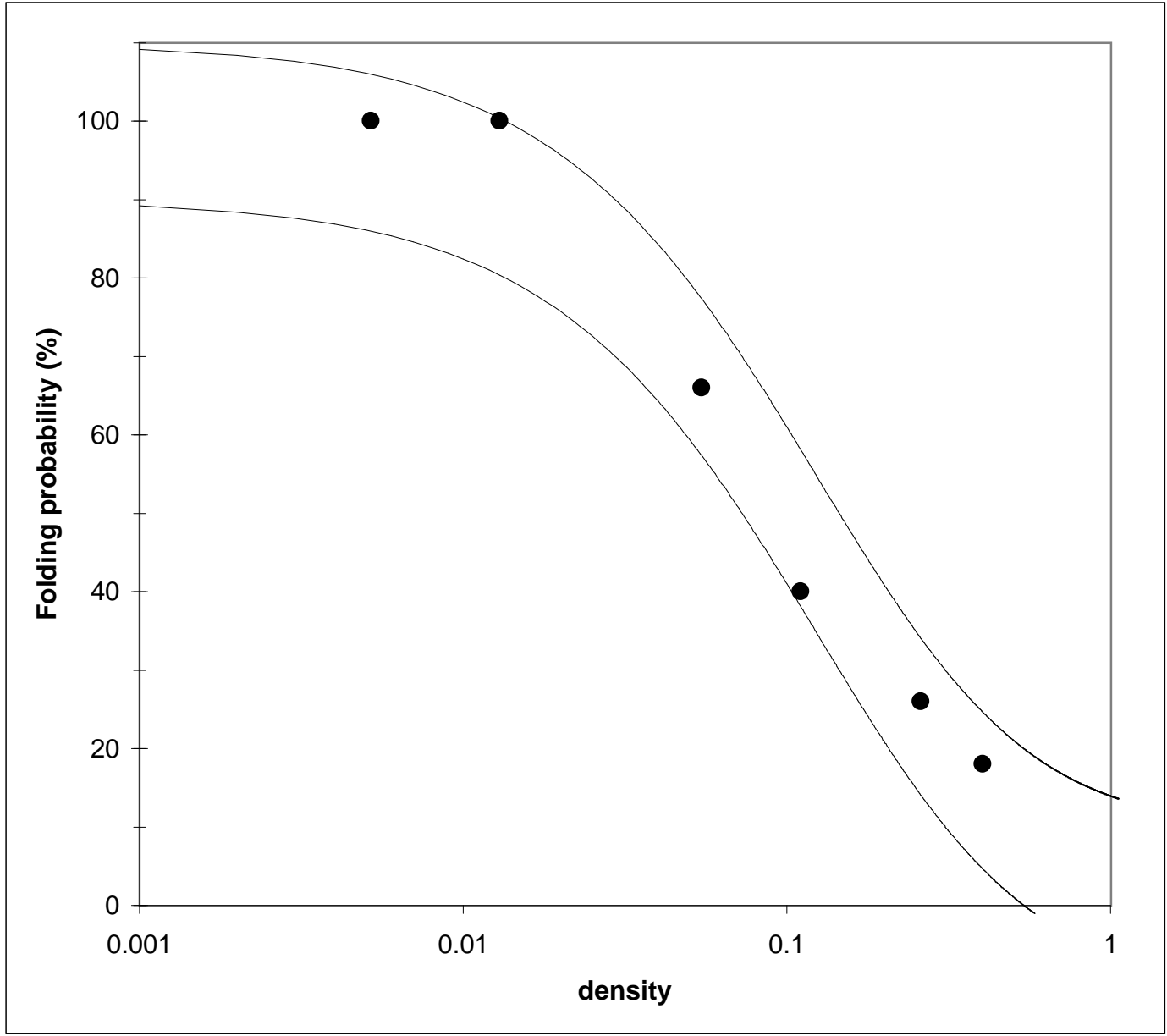
\end{document}